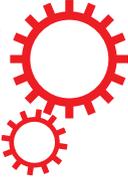

# The onset of dissipation in high-temperature superconductors: magnetic hysteresis and field dependence

E. F. Talantsev[1,2], N. M. Strickland[3], S. C. Wimbush[3,4], J. Brooks[3], A. E. Pantoja[3], R. A. Badcock[3], J. G. Storey[3,4] & J. L. Tallon[3,4]

Recently, we showed that the self-field transport critical current, $I_c$(sf), of a superconducting wire can be defined in a more fundamental way than the conventional (and arbitrary) electric field criterion, $E_c = 1\,\mu$V/cm. We defined $I_c$(sf) as the threshold current, $I_{c,B}$, at which the perpendicular component of the local magnetic flux density, $B_\perp$, measured at any point on the surface of a high-temperature superconducting tape abruptly crosses over from a non-linear to a linear dependence with increasing transport current. This effect results from the current distribution across the tape width progressively transitioning from non-uniform to uniform. The completion of this progressive transition was found to be singular. It coincides with the first discernible onset of dissipation and immediately precedes the formation of a measureable electric field. Here, we show that the same $I_{c,B}$ definition of critical currents applies in the presence of an external applied magnetic field, $B_a$. In all experimental data presented here $I_{c,B}$ is found to be significantly (10–30%) lower than $I_{c,E}$ determined by the common electric field criterion of $E_c = 1\,\mu$V/cm, and $E_c$ to be up to 50 times lower at $I_{c,B}$ than at $I_{c,E}$.

Dissipation-free electric current transport is one of the most fascinating and practically important properties of superconductors. The fundamental origins of the limitations to these dissipation-free currents, including conditions in which an external magnetic field, $B_a$, is applied, are of central interest in both pure and applied superconductivity.

The conventional approach to describing dissipation-free currents in superconductors is based on the concept of a critical current, $I_c$, which is commonly defined as the current at which the electric field, $E$, along the conductor resulting from the onset of dissipation reaches a certain measurable value, $E_c$[1], usually defined to be $1\,\mu$V/cm[2,3]. We will designate critical currents defined in this way by $I_{c,E}$. There have been several proposals to define the critical current based on other criteria, for instance, a resistive criterion or a power dissipation per unit volume criterion[4]. The complexity of the problem was discussed in detail in many reports[4–6].

Despite the fact that these various definitions of $I_c$ satisfy the requirements of many practical applications of superconductors, they do not however reflect the true nature of the limits to dissipation-free current flow in superconductors for which the resulting electric field $E$, as well as the resistance and power dissipation should be zero.

In attempting to resolve this problem, recently[7,8] we showed that the definition of the dissipation-free critical current under self-field conditions, $I_c$(sf) (when no external magnetic field is applied) can be arrived at in a more fundamental way, based not on a continuous onset of dissipation with an arbitrary threshold, but on an abrupt physical effect which we identified in high-temperature superconducting (HTS) tapes by measuring and analyzing the perpendicular component of the magnetic flux density, $B_\perp(x)$, generated across the surface of the conductor by the transport current. Here $x$ is the position across the surface of the conductor of width $2a$ running from $x = -a$ to $x = +a$. We observed a distinct threshold value of the transport current at which dissipation sets in. More importantly, at this threshold current, $I_{c,B}$, three simultaneous processes occur:

1. An abrupt transition from a non-linear to a linear dependence of $B_\perp(I)$, as observed at all locations, $x$, on the superconductor surface.
2. A crossover from a non-uniform to a uniform current distribution across the tape width[7–10].
3. The abrupt onset of a non-zero electric field along the wire. This implies that $I_{c,B}$ is the current at which dissipation commences.

[1]M.N. Mikheev Institute of Metal Physics, Ural Branch, Russian Academy of Sciences, 18, S. Kovalevskoy St, Ekaterinburg, 620108, Russia. [2]NANOTECH Centre, Ural Federal University, 19 Mira St., Ekaterinburg, 620002, Russia. [3]Robinson Research Institute, Victoria University of Wellington, 69 Gracefield Road, Lower Hutt, 5010, New Zealand. [4]The MacDiarmid Institute for Advanced Materials and Nanotechnology, Victoria University of Wellington, P O Box 600, Wellington, 6140, New Zealand. Correspondence and requests for materials should be addressed to E.F.T. (email: evgeny.f.talantsev@gmail.com)

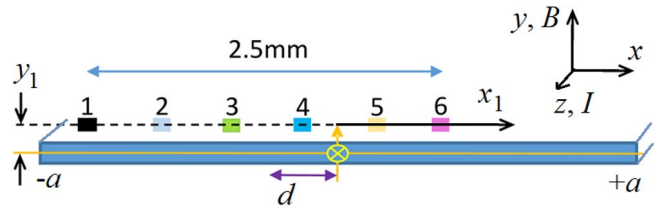

**Figure 1.** Schematic representation of the experiment to study $B_{perp}(I)$ of HTS tapes. The violet arrow shows the distance, $d$, from the centre of the Hall sensor array to the centre of the wire for the 6-sensor array in which the sensors were spaced by 0.5 mm, with a total array width of 2.5 mm. $B_\perp$ is measured and calculated at coordinates $(x_1, y_1)$ and the axes show the $x$, $y$ and $z$ directions as well as the current, $I$, direction along the tape and the external applied field, $B$, direction normal to the tape.

The observed non-linear behaviour arises from current redistribution across the tape width as the critical state progresses from the edges towards the centre; and the abrupt transition to linear behaviour occurs when the entire conductor divergently reaches the critical state, as characterised by a uniform current. It is tempting to infer that, because $B_\perp(I, x)$ suddenly becomes linear in $I$ at *all* points across the surface then this implies a transformation of the *whole* tape at $I_{c,B}$. However, $B_\perp(I, x)$ reflects the local current density *integrated across the entire width*. Just below $I_{c,B}$ much of the tape is already in the critical state and it is simply the last singular transition to the critical state near the centre that causes a discontinuity in $B_\perp(I, x)$ at all points $x$ across the surface. Nonetheless the critical state is thermodynamically determined and thus $I_c$ has a thermodynamic origin[7], rather than one rooted in the initiation of localised dissipation hotspots in some weak-linked or high current density areas. Because of this we propose to designate $I_{c,B}$ as the fundamental critical current, one which is free from arbitrary criterion-based definitions such as customized electric field, resistance or power-dissipation criteria.

Others have used Hall probes to measure the field distribution across HTS conductors. For example, Tallouli and coworkers[11,12], and references therein measured the perpendicular field profile over $(Bi, Pb)_2Sr_2Ca_2Cu_3O_{10}$ and $YBa_2Cu_3O_y$ tapes and deconvolved this ("the inverse problem") to infer the differing current profiles for fast and slow transient excitation.

Our focus is to describe the spatial evolution of the critical state and demonstrate the fundamental thermodynamic origins of the onset of dissipation in practical conductors. We show that our definition of self-field critical currents as $I_{c,B}(sf)$ is naturally extended to in-field critical currents, $I_{c,B}(B)$. We also show that the hysteretic evolution of the current and field distribution can be accurately calculated using the equations of Norris[13], Brandt and Indenbom[14] and Zeldov et al.[15] for both self-field (i.e. zero externally-applied field) and in-field measurements.

## Experiment

Two distinct experiments were performed on commercial $REBa_2Cu_3O_7$ (REBCO) coated conductor tapes (second-generation, 2 G, wire). In both experiments the applied field is normal to the tape surface i.e. parallel to the crystallographic c-axis. The first experiment, carried out using a 1.5-Tesla-scale laboratory magnet/cryostat, studied tapes manufactured by SuNAM Co., Ltd. (SAN04200-161031-01) having a width of $2a = 4$ mm and a GdBCO layer thickness of $2b = 1.9 \mu m$. Angular in-field critical current data, $I_{c,E}(T, B, \theta)$ for this wire is also available[16]. Measurements were performed with the samples immersed in a liquid nitrogen cryostat.

A cryogenic Hall-effect sensor array (MULTI-7U) obtained from Arepoc s.r.o. (Slovakia) was used to measure the perpendicular component of the local surface magnetic flux density, $B_\perp$, at various points across the width of the sample as shown schematically in Fig. 1. This type of sensor is used in many reports[7–10,17,18]. Each individual sensor has an active area of 0.01 mm² (100 μm × 100 μm). We used six sensors of this array spaced 0.5 mm apart, giving a total array width of 2.5 mm. We designate the sensors by the numbers 1, 2, 3, 4, 5 and 6. The sensor array is mounted face down beneath the HTS tape separated by a distance of 0.85 mm to protect the exposed sensor elements. This is the distance, $y_1$, shown in Fig. 1. In our analysis below we confirm and refine this estimate of $y_1$ by comparing measurements and calculations of $B_\perp(x)$.

For practical wires and cables transport current data are often presented in the form of $I$-$E$ curves[19]. We used the same approach here. To deduce $I_{c,E}$ values, experimental $I$-$E$ datasets were fitted to following equation:

$$E(I) = E_0 + k \cdot I + E_c \cdot \left(\frac{I}{I_{c,E}}\right)^n \quad (1)$$

where $E_0$ is an instrumental offset and $k$ is a linear term used to accommodate incomplete current transfer in short samples.

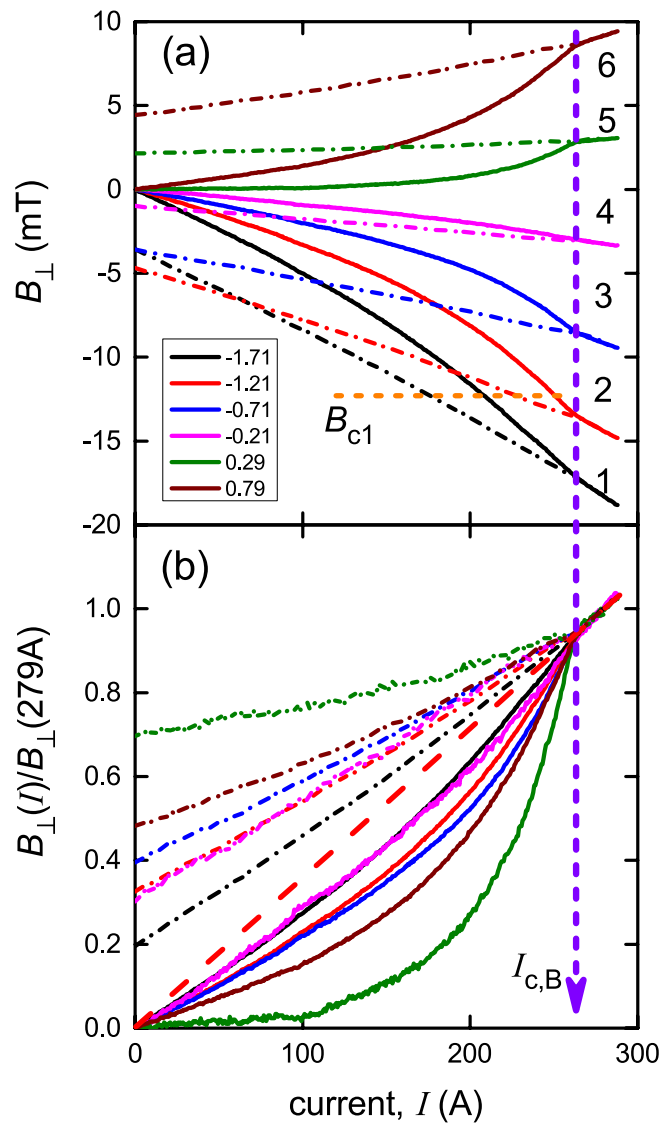

**Figure 2.** Experimental values of (**a**) $B_\perp(I)$ and (**b**) $B_\perp(I)/B_\perp(I = 279\,\text{A})$ for a transport current ramp up to 280 A (solid curves) and back to zero current (dash-dot curves) through a 4 mm wide SuNAM 2 G wire under self-field conditions. The $x_1$ locations of the sensors are indicated in mm in the legend. The dashed red line in panel (**b**) shows the extrapolation of the linear part of $B_\perp(I)/B_\perp(I = 279\,\text{A})$ back to $I = 0$ A. In panel (**a**) the magnitude of $B_{c1} = 12.3 \pm 0.1$ mT is indicated by the horizontal orange dashed line.

In the second experiment using a different apparatus we measured the field distribution across the width of 10 mm wide GdBCO coated-conductor tapes manufactured by Fujikura Ltd. (FYSC-S10 10-0025-02). The samples were immersed in a bath of liquid nitrogen (i.e. at 77 K) and a Helmholtz coil was used to apply a field up to $B_a = 80$ mT (data not reported below). The voltage taps for measuring critical current were positioned along the tapes approximately 12 cm apart, with a Hall sensor array placed half way between the two voltage taps, spanning the width of the conductor. Each sample length of superconducting tape used was 15 to 20 cm long between the current supply leads. Voltage measurements were acquired on an Agilent 34420 A nanovoltmeter using a 100 μV range, and the transport current was supplied by an Agilent 6680 A constant-current power supply. $V$-$I$ characteristics were measured up to 600 A and the critical current was, again, determined according to the usual 1 μV/cm electric-field criterion.

In these experiments the Hall sensor array is a seven-element linear array (THV-MOD) also manufactured by Arepoc s.r.o. (Slovakia) operating with an excitation current of 4 mA. The manufacturer's field and temperature sensitivity calibrations were used for each individual sensor. A seven-channel preamplifier with a DC gain of 1300 was used to amplify the Hall voltage signals, and the data was captured with a National Instruments c-DAQ acquisition system with an integration time of 0.5 s, run under a LabVIEW platform. The absolute accuracy of each sensor was verified using an N38 Nd-Fe-B magnet and a calibrated Hall probe (Group 3, type LPT-141), and found to be accurate to better than 2%. The sensitivity of the system is ≤0.02 mT, and each sensor has a specified linearity better than 0.2% up to 1 T. In Figs 3, 6 and 9, below, the errors in the $B_\perp$ data points are smaller than the size of the data points and thus have been omitted. The Hall sensors each have an active area of 0.05 mm², are linearly positioned 1.5 mm apart and lie approximately 0.5 mm above the superconductor. The sensor array was positioned both centred on the tape and off-centre so as to scan the tape edges where the field profile changes most rapidly. We found that centred and off-centre measurements overlaid each other quite precisely.

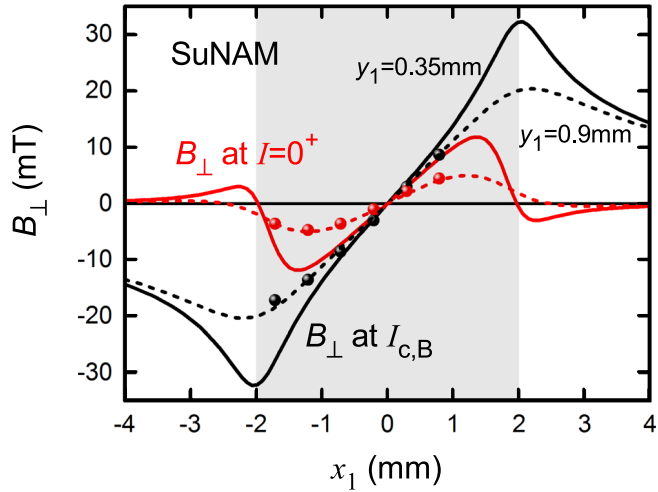

**Figure 3.** Experimental and calculated values of $B_\perp(x_1)$ across the width of the 4 mm wide SuNAM tape when (i) the current has been raised to $I_{c,B} = 264$ A (black curves and data points) then (ii) reduced back to $I = 0^+$ (red curves and data points). The shaded region shows the extent of the conductor width. Solid (dashed) calculated curves are for sensors located at 0.35 mm (0.9 mm) above the superconducting film. Evidently the agreement is best for $y_1 = 0.9$ mm which we then use in the calculations for $B_\perp(I)$ shown in Fig. 4.

**Results and Discussion**

As superconductors are highly hysteretic magnetic materials[20], any prior transport current flow, which generates a self-magnetic field, affects the subsequent $B_\perp(I)$ experiment, because the self-generated magnetic field will always (to a greater or lesser extent) be trapped by the superconductor, thereby persisting even after the current flow ceases. To the authors' knowledge, the effect of trapping of the self-magnetic field by the superconductor under the conditions of transport current flow has not previously been studied experimentally.

Here we report our experimental findings associated with hysteretic flux trapping under conditions of transport current flow. We also present calculations for these effects and show that all results under self-field are essentially well understood and accurately described by these calculations.

**Self-field conditions.** *Self-field measurements – SuNAM.* In Figs 2 and 3 we present experimental results for SuNAM 2 G wire for which the Hall sensor array was centred at the position $d = -0.5$ mm. In this position all six Hall sensors are located inside the tape width, with sensor 1 located at a distance of 0.25 mm from the left edge of the wire. In Section 3.1.2 we locate the sensor positions more accurately using the fact that $B_\perp(x)$ is an odd function of $x$. Electric potential taps (which were used to measure the induced electric field) were soldered to the wire at a separation of 48 mm along the tape. The conventionally defined critical current for this sample was found to be $I_{c,E} = 279$ A with an $n$-value of 35.

The wire was first zero-field and zero-current cooled and then subjected to a transport current ramp rising from 0 to 288 A, after which the current was decreased back to 0 A. We refer to this final state as $I = 0^+$.

Three important experimental findings can be immediately reported:

1. *Hysteresis*: for all six Hall sensors, the starting and end points for surface flux density, $B_\perp(I=0)$, are different. After the first current flow, magnetic flux was trapped differently by different parts of the tape. Notably, $B_\perp(I)$ curves on reducing current are more linear in comparison with the virgin ramp.
2. *Linearity above $I_{c,B}$*: For each Hall sensor, $B_\perp(I)$ curves for both rising and falling current collapse onto the same straight line when the transport current exceeds $I_{c,B} = 264 \pm 3$ A. As we already showed in[7,8], linear extrapolation of this straight line back to $I = 0$ accurately intersects the vertical axis at $B = 0$ T – see dashed red line in Fig. 2(b). Such linearity of $B_\perp(I)$ is a common characteristic of the resistive state[21].
3. *Penetration depth*: the measured value of $I_{c,B}(77\,K) = 264 \pm 3$ A can be used to calculate the critical current density, $J_{c,B}(77\,K)$, which can be used to deduce the London penetration depth, $\lambda_{ab}(77\,K)$, for this wire by numerical solution of the basic equation[22] in terms of $\lambda_{ab}$:

$$J_c(sf) = \frac{\varphi_0}{4\pi\mu_0} \cdot \frac{\ln(\kappa) + 0.5}{\lambda_{ab}^3} \cdot \left[\frac{\lambda_c}{b} \cdot \tanh\left(\frac{b}{\lambda_c}\right)\right] \quad (2)$$

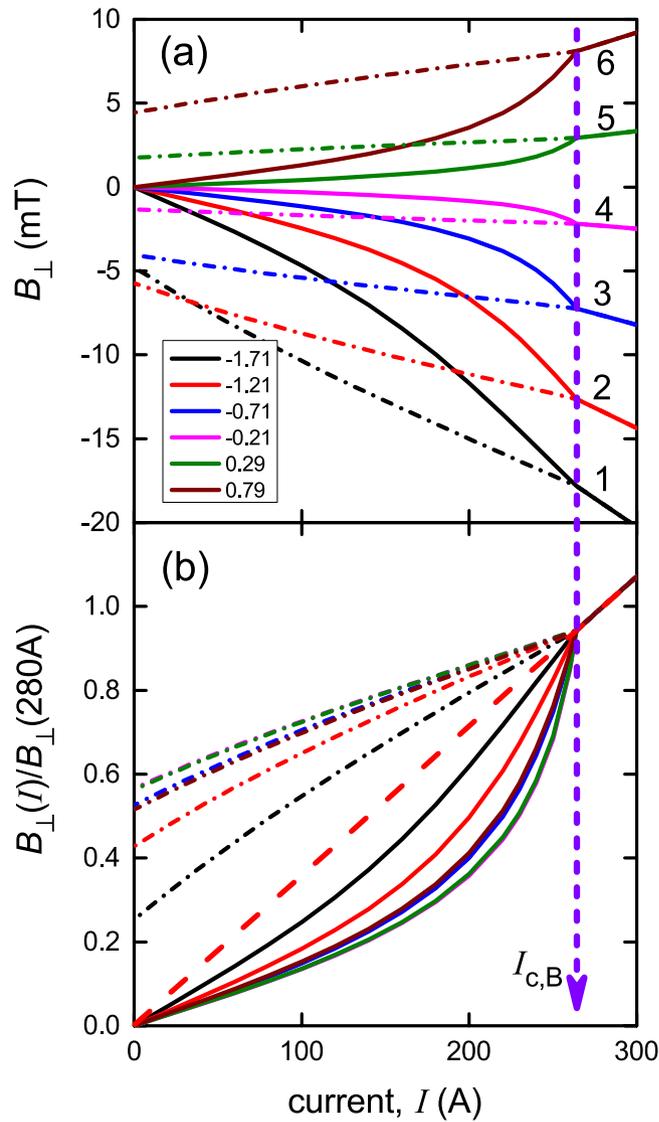

**Figure 4.** Calculated values of $B_\perp$ as a function of current, $I$, for the six sensors located at $x_1 = -1.71$, $-1.21$, $-0.71$, $-0.21$, $+0.29$ and $+0.79$ mm across the width of the 4 mm wide SuNAM tape. Currents increasing to 300 A are shown by solid curves and currents then decreasing to $0^+$ are shown by dash-dot curves. The red dashed line shows the extrapolation back to $I = 0$ of the linear behaviour seen above $I_{c,B} = 264$ A. The comparison with the experimental data shown in Fig. 2 is excellent. $I_{c,B}$ is thus the termination of the non-linear $B_\perp(I)$ behaviour and is the true thermodynamic critical current. The calculations are carried out for $y_1 = 0.9$ mm.

where $\kappa = \lambda/\xi = 95$ for YBCO[20], and the term in square brackets is the thickness correction factor for anisotropic superconductors[22,23]. The electron mass anisotropy, $\gamma = \lambda_c/\lambda_{ab}$, can be taken to be equal to 7 which is a typical value over most representative experimental data for $YBa_2Cu_3O_7$[9] and is in fact the value we previously found using the geometrical form factor in Eq. (2) using many different thicknesses, $2b$. The resulting value of $\lambda_{ab}(77\,K) = 260 \pm 1$ nm was used to determine the lower critical field, $B_{c1}(77\,K)$, of this SuNAM wire as[22,23]:

$$B_{c1}(77\,K) = \frac{\varphi_0}{4\pi} \cdot \frac{\ln(\kappa) + 0.5}{\lambda_{ab}^2(T = 77.4K)} = 12.3 \pm 0.1 \text{ mT} \qquad (3)$$

Note that the reduced error in $\lambda$ relative to the (small) error in $J_c$ arises from the cube root that results from inverting Eq. (2). The deduced value of $B_{c1} = 12.3 \pm 0.1$ mT is in good agreement with the value reported for $YBa_2Cu_3O_7$ single crystals[24] for $B$ applied normal to the $CuO_2$ planes. The physical meaning of $B_{c1}$ is that it separates, at equilibrium, the pure Meissner state of the superconductor (for $B < B_{c1}$), under which the magnetic field penetrates to a depth of only a few $\lambda$, from the mixed state ($B_{c1} < B < B_{c2}$) under which the magnetic field penetrates much deeper in the form of Abrikosov vortices, where the local magnetic field near the vortex cores is much greater than the average field. Thus, in Fig. 2 we indicate the $B_{c1}$ boundary which in principle separates the two magnetically different states of the superconductor, but in fact there is no apparent anomalous effect on $B_\perp$ at either of the two locations (sensors 1 and 2) where $B_\perp$ exceeds $B_{c1}$. For example, $B_\perp$ for sensor 1 reaches $B_{c1}$ at $I = 215$ A on the virgin ramp but $B_\perp(I)$ remains featureless until $I = I_{c,B} = 264$ A, while, for sensor 2, $B_\perp$ reaches $B_{c1}$ at $I = 250$ A but again $B_\perp(I)$ remains featureless until $I = I_{c,B} = 264$ A where dissipation first sets in. We conclude that the transition from dissipation-free current flow to the dissipative regime is not related to a possible crossover from the Meissner state into the mixed state. Moreover, we previously demonstrated[9] for thicker films and single crystals that $J_c$ closely followed the London geometrical form factor in the square brackets of Eq. (2) – a further indication that these samples remain in the Meissner state in self-field.

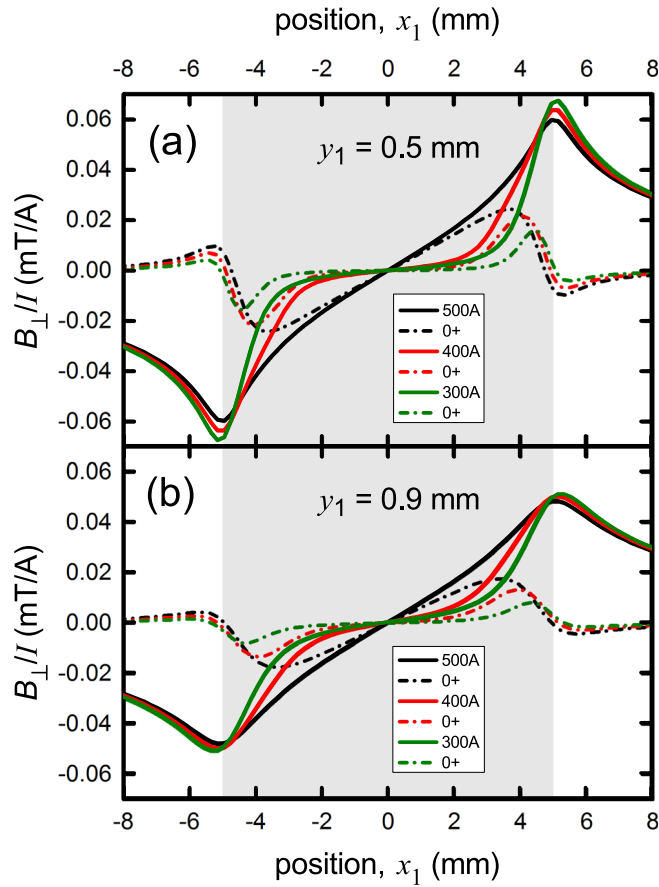

**Figure 5.** The calculated perpendicular field distribution, $B_\perp(x_1)$, across the width of a 10 mm wide Fujikura superconducting 2 G tape where the current is raised to 500 A (solid black), 400 A (solid red) and 300 A (solid green) then, for each, returned to zero transport current (accompanying dash-dot curves). The shaded region shows the extent of the conductor width. Two cases are shown where the sensor array and superconducting film are separated by (**a**) 0.5 mm and (**b**) 0.9 mm. The effect of moving the sensors further away is to round out the sharper features seen at the conductor edges.

*Self-field calculations – SuNAM.* We use the approach detailed in Brandt and Indenbom[14] for calculating the self-field current distribution, in particular their Eqs (2.4), (2.5), (3.5), (3.6) and (3.7). As before[10] the distribution of the perpendicular component of the field $B_\perp$ at the point $(x_1, y_1)$ above the conductor is calculated from Ampere's law using:

$$B_\perp(x_1, y_1) = \frac{\mu_0}{2\pi} \int_{-a}^{+a} j(x, I) \frac{x - x_1}{(x - x_1)^2 + y_1^2} dx \qquad (4)$$

where $j(x, I)$ is the local $x$-dependent current density integrated over the film thickness, given by the above-noted equations from Brandt and Indenbom. As noted above, the coordinate $y_1$ is the effective height of the active sensor array above the superconducting film while $x_1$ is the location, in the plane of the sensors, along the $x$-axis with origin at the centre. Figure 3 shows the calculated distribution of $B_\perp(x_1)$ when the current is raised to $I_{c,B} = 264$ A (black curves) and then reduced back to $I = 0^+$. We use the two values of $y_1$, namely $y_1 = 0.35$ mm, the physical separation of the sensor body and the HTS tape and $y_1 = 0.9$ mm, the value which gives a best fit, here and in the following. The accompanying data points are the experimentally observed values taken from Fig. 2.

The double inflexion of the red curve at the tape edges reflects the trapped flux near the centre of the film and the effect of associated screening currents which for $I = 0^+$ sum to zero i.e. negative currents flow at the edges. The effect of increasing the value of $y_1$ is to broaden the sharp features seen at the tape edges. Evidently the best fit to the experimental data is for the value of $y_1 = 0.9$ mm which is close to the actual physical separation of sensor and tape, $y_1 = 0.85$ mm, and this value is consistent with the inverted sensor geometry as noted in Section II. Because the calculations are rather sensitive to the value of $y_1$ we believe this value of 0.9 mm is an accurate measure of the true position of the active sensor region relative to the superconducting film.

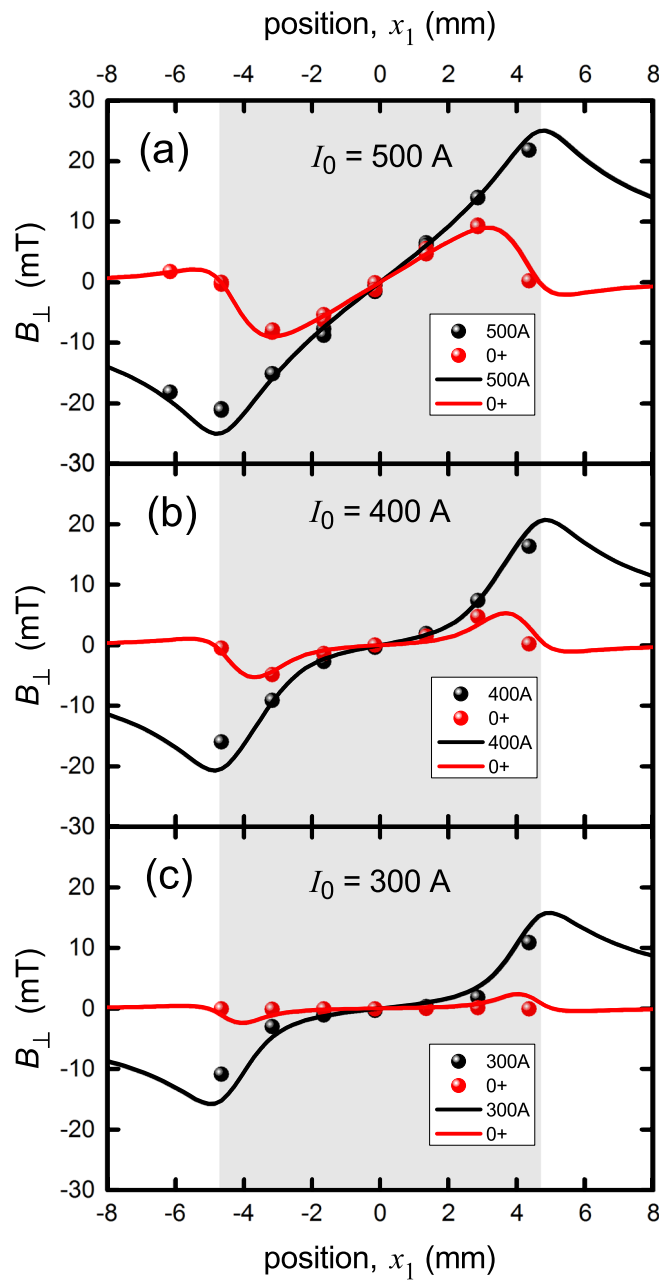

**Figure 6.** Comparison between the measured and calculated field distribution, $B_\perp(x_1)$, across the width of the nominally 10 mm wide Fujikura superconducting 2 G tape when the current is raised to $I_0$ (black curve) then reduced to zero (red curve). Panels (a), (b) and (c) are for $I_0 = 500$ A, 400 A and 300 A, respectively. The calculations are for width $2a = 9.4$ mm and $y_1 = 0.9$ mm. The shaded region shows the extent of the conductor width, as used in the calculations. The data and calculations correspond well and show all the same general trends. The small differences at the edges could be attributable to substrate susceptibility which tends to smooth out the larger variations.

The $x$-dependent distribution of $B_\perp(I)$, being an odd function of $x$, allows us to accurately locate the $x$-position of the sensor array relative to the film centre. In this way we found that the array was an additional 0.04 mm to the right and the sensor positions listed in the legend of Fig. 2 were obtained in this way. Having established the precise location of the sensor array from Fig. 3 we proceed to calculate the full current dependence of $B_\perp$ for each sensor location. The results are shown in Fig. 4 and the comparison with the experimental data in Fig. 2 is very good. The only differences can be seen in panel (b) for sensors 4 and 5 near the centre where the absolute magnitude of $B_\perp$ is small and the rescaling amplifies any minor differences.

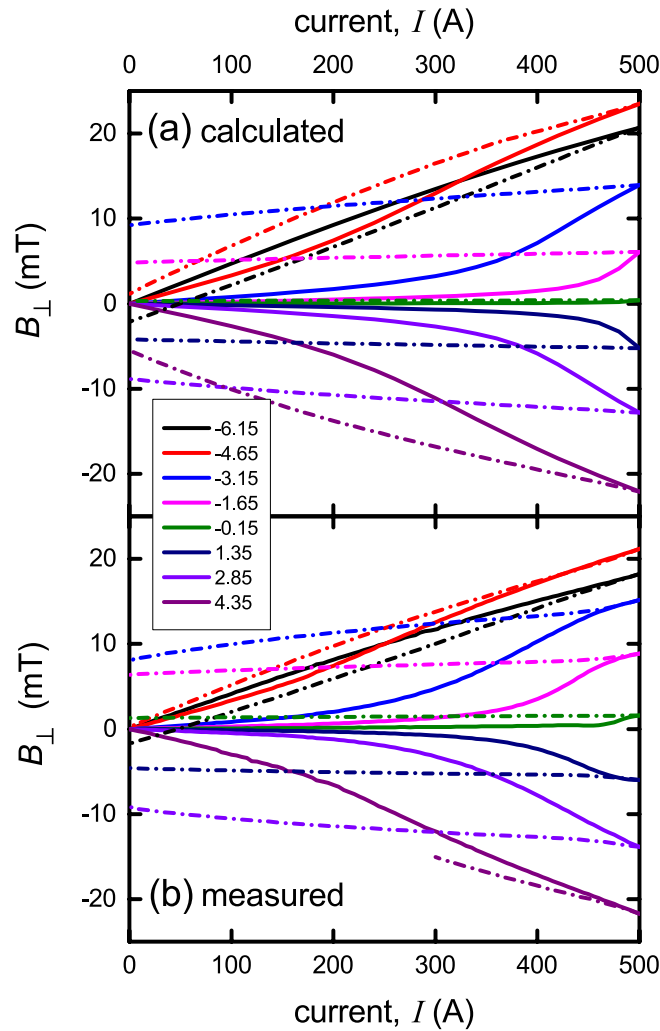

**Figure 7.** Comparison between (**a**) the calculated and (**b**) measured field distribution, $B_\perp(x_1)$, across the width of a 10 mm wide Fujikura superconducting 2 G tape when current is raised to $I_0 = 500$ A and returned to zero. Each of the curves corresponds to the Hall sensor $x_1$ location as listed in the figure legend. Note that all sensors show a rise from zero in the virgin curves but after reaching $I_0$ return to finite values reflecting the local inversion of the critical state, along with currents and counter shielding currents associated with trapped flux. The sensor at $+4.35$ mm failed on cycling back to zero (purple curve).

This then confirms that the evolution of the current distribution is associated with the onset of a critical state at the edges which progressively moves inwards to the centre as the current is increased thus resulting in the non-linear evolution of $B_\perp(I)$. The critical state abruptly reaches the centre at $I_{c,B}$ coinciding with the first onset of observable dissipation and thereafter $B_\perp(I)$ reverts to linear behaviour. We refer to $I_{c,B}$ as the thermodynamic or "true" critical current[7]. This critical state, wherever its location on the conductor, lies at the threshold of dissipation and it is important to note that these experiments in themselves do not identify the microscopic origin of the dissipation, whether it is vortex depinning[14], depairing of Cooper pairs[25], spin currents[26], or based on quantum limitation[27]. Despite this, the critical-state model seems to give a full description of the current and field distribution, as will be further confirmed below, and it remains only to quantify the effect of the paramagnetic Hastelloy substrate. This is a future objective.

It is interesting to note that these calculations show that this non-linear $B_\perp(I)$ behaviour prior to the onset of dissipation is purely a consequence of the rectangular cross-sectional geometry of the superconducting film. This leads to the prediction that such a transition will not be observed in cylindrical symmetry. Instead, for a uniform sample, the critical state is reached simultaneously around the entire circumference and from Ampere's law the surface azimuthal field will always remain linear in $I$ whether or not the sample is in the superconducting, critical, mixed or normal states.

*Self-field measurements and calculations – Fujikura.* For an ideal tape conductor, where the superconducting film has a rectangular cross-section, the only adjustable parameter in our calculations is the coordinate, $y_1$, of the plane of the Hall sensors relative to the superconducting film. In this section the calculations are made using Eq. (2.11) of Brandt and Indenbom[14]. Figure 5 shows $B_\perp(x_1)$ for the case of our 10 mm wide Fujikura tape with a surface-field critical current density, $I_{c,B}$, equal to 500 A, as observed. As noted, this is the critical current at which the critical state just extends across the entire tape and $B_\perp(x_1, I)$ abruptly crosses over from non-linear to linear. In the figure we show the calculated field distribution when the current is raised (in the virgin state) from zero to $I_0 = 500$ A, 400 A and 300 A (solid curves) and then lowered back to zero for each of these (dash-dot curves). The top panel (a) is for $y_1 = 0.5$ mm (the hypothetical separation of the sensor and tape) and the lower (b) is for $y_1 = 0.9$ mm (our best estimate of the effective separation of tape and the active region of the sensor). The overall behavior is similar but the effect of displacing the Hall array more distant from the conductor is to round out, and weaken, the more abrupt changes that occur at the film edges.

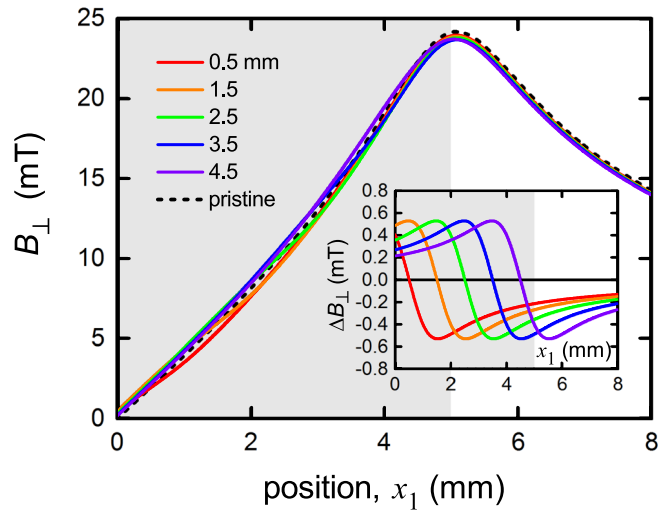

**Figure 8.** Calculated perpendicular flux density, $B_\perp(x_1)$, for a 10 mm wide Fujikura 2 G wire at $I_{c,B}$ when a 1 mm wide section has a 10% local reduction in $J_c$. Five such locations are shown: $x_1 = 0.5$ mm, 1.5 mm, 2.5 mm, 3.5 mm and 4.5 mm (edge), and are annotated as such. The dashed curve shows the pristine response and the inset shows the modulations, $\Delta B_\perp$, about the pristine response. Shading shows the extent of the conductor in the $x_1$ direction, as used in the calculation. The size of these modulations, exceeding 1.1 mT, are easily detectable by the Hall sensor array.

Figure 6 shows a comparison of calculations with experimental data for the nominally 10 mm wide Fujikura wire where we used $y_1 = 0.9$ mm. It was evident from the small misfit at the edges that the active superconducting width might not be the full 10 mm width but more like 9.4 mm and the calculations were performed for this width. Though the misfit was small, these measurements are very sensitive to attenuation of superconducting properties near the edges or indeed anywhere over the cross-section. It is also possible that the quite significant paramagnetic susceptibility found for Hastelloy at 77 K[28] plays a role in rounding out the more abrupt features of the field profile at the edges[29].

Despite these small differences all the overall generic features as they evolve with current are present including, most notably, the change of sign in $B_\perp$ at $x_1 = -6$ mm when the current has been reduced back to $0^+$ (see panel (a) red curve and data point). This is a consequence of the screening currents associated with trapped flux together with the fact that the total transport current sums to zero. Thus there remains a positive current across the centre and counterbalancing negative currents at the two edges which give rise to the small negative $B_\perp$ there. The measurements and calculations also show that there is very little residual field after raising the current to 300 A then returning to $I = 0^+$, even though this current is a full two thirds of $I_{c,B}$. This very non-linear response of the residual field arises from the fact that in panel 6(a), where the current has been raised to $I_{c,B}$, on reducing $I$ to $0^+$ we find positive current remains across much of the cross section and only at the edges is the current reduced to negative $I_{c,B}$. In contrast, in panel (c) the saturation region extends only a certain distance in from the edges and then, on reducing the transport current to $0^+$, the positive and negative screening currents (which sum to zero) are confined more to the edges and substantially cancel in the calculation of $B_\perp$.

Note that when $I_0 = 500$ A, the saturated critical state extends all the way to the central axis of the tape, i.e. the current distribution is uniform across the width of the tape. When $I_0 = 400$ A it extends in some 20% of the total width from either edge, and when $I_0 = 300$ A, the critical region extends in only 10% from either side. In general, the uniform critical region extends inwards a distance $d$ given by ([14] Eq. 2.5):

$$d(I) = a \times \sqrt{1 - (I/I_c)^2} \qquad (5)$$

One should also note that the edge quality on one side may be better or poorer than the other side. There is evidence of such performance asymmetry (though slight) in Fig. 6. We note however that our measurements show that these are excellent quality superconducting films from the perspective of current uniformity. This will be explored in more detail below.

Figure 7 shows the current dependence of $B_\perp$ for each of the sensors located at $x_1 = -6.15, -4.65, -3.15, -1.65, -0.15, +1.35, +2.85$ and $+4.35$ mm. Panel (a) shows the calculated data while panel (b) shows the experimentally-observed data. The match is good. Below 500 A, the dependence of $B_\perp(I)$ is highly non-linear while for $I > I_0 = 500$ A the dependence is immediately linear (not shown). This distinguishes $I_0 = 500$ A as the thermodynamic critical current.

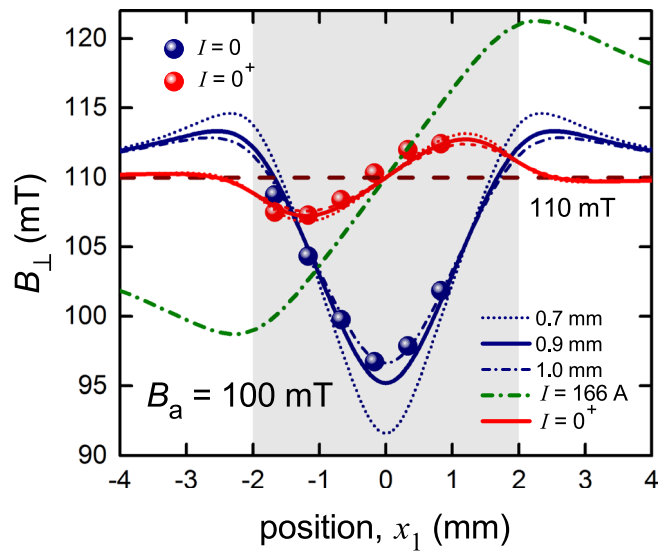

**Figure 9.** Surface perpendicular flux density, $B_\perp(x_1)$, for 4 mm wide SuNAM 2 G wire ($d = -0.5$ mm) at two stages illustrating hysteresis: (i) on application of $B_{appl} = 100$ mT to the zero-field-cooled sample (blue – denoted $I = 0$); and (ii) after a transport current with amplitude of 207 A was applied then reduced back to zero (red – denoted $I = 0^+$). The shaded region shows the extent of the conductor width. While the applied field is 100 mT the local field is enhanced to $110 \pm 0.5$ mT due to the susceptibility of the Hastelloy substrate and as a consequence the data is distributed about this value. The blue curves are the calculated values of $B_\perp$ for $y_1 = 0.7$, 0.9 and 1.0 mm for an applied field of 110 mT (the substrate is ignored in the calculations). The olive dash-dot curve shows the calculated $B_\perp$ profile for $B_a = 110$ mT and $I = I_{c,B} = 166$ A. The red curves are for $I$ raised to 166 A then reduced to $0^+$ and the dashed and dotted curves correspond to different $y_1$ values as indicated in the legend.

Bearing in mind that the measurement sensitivity is far better than 0.5 mT the dominant cause of deviations in the systematic measurements presented as a function of $x_1$-coordinate is the conductor itself – the inevitable variations in local $J_c$ that one seeks to minimise but which our measurements are especially sensitive to. As noted, these conductors are of excellent uniform performance but our measurements are capable of exposing rather small modulations in current density. We now explore this question in a little more detail.

Figure 8 shows the effect on the field distribution of a local 10% reduction in $J_c$ over a section of conductor of 1 mm width located at various points across the surface. These are $x_1 = 0.5$ mm, 1.5 mm, 2.5 mm, 3.5 mm and 4.5 mm (the edge), as annotated. The dashed curve shows the pristine response for a uniform conductor. The modulations, $\Delta B_\perp$, relative to the pristine response are shown in expanded view in the inset and their amplitude exceeds 1.1 mT – a figure easily detectable by our Hall sensor array. In light of these comments the agreement between measurements and calculation are exceptionally good in terms of the overall generic evolution of local current density as it is cycled up and down.

**In-field conditions.** *In-field measurements – SuNAM.* To demonstrate the flux trapping/magnetic hysteresis effect under the conditions where an external magnetic field, $B_a$, is applied perpendicularly to the tape, we measured the same SuNAM 2 G wire sample reported in Sections 3.1.1 and 3.1.2. The specimen was zero-field cooled and then an external magnetic field of $B_a = 100$ mT was applied (Fig. 8, blue data points). This value was chosen such that the entire sample could, in principle, be in the mixed state, as $B_a \gg B_{c1}(77\,K) = 12.3$ mT. As expected[14,15], the $B_\perp(x_1)$ profile exhibits a "V" shape centred on the middle of the wire.

Now, when a transport current with amplitude of $I = 207$ A, in excess of both $I_{c,B}$ and $I_{c,E}$ at this field (Fig. 4), is applied to the sample and then reduced to zero while still maintaining $B_a = 100$ mT, the trapped $B_\perp$ values appear to move upwards and develop an asymmetry about the tape centre. This lift and asymmetry is in no way anomalous as the accompanying calculations show.

We may use the equations developed by Zeldov *et al.*[15] but in fact the applied field and current are sufficiently high that they lead to current saturation across the conductor, so here we may integrate Eq. (4) analytically. Firstly we anticipate the result found below that at 100 mT, $I_{c,B}$ is reduced to 166 A. For the case of zero applied current in a 100 mT applied field the current is saturated at $-166$ A for the left half of the conductor and at $+166$ A for the right half. In the region $-b \leq x \leq +b$ the current crosses from negative to positive according to $166 \times (2/\pi) \times$ arc-sin$(x/b)$. Provided $b \ll a$, this contribution to the integral (Eq. (4)) can be neglected and we obtain:

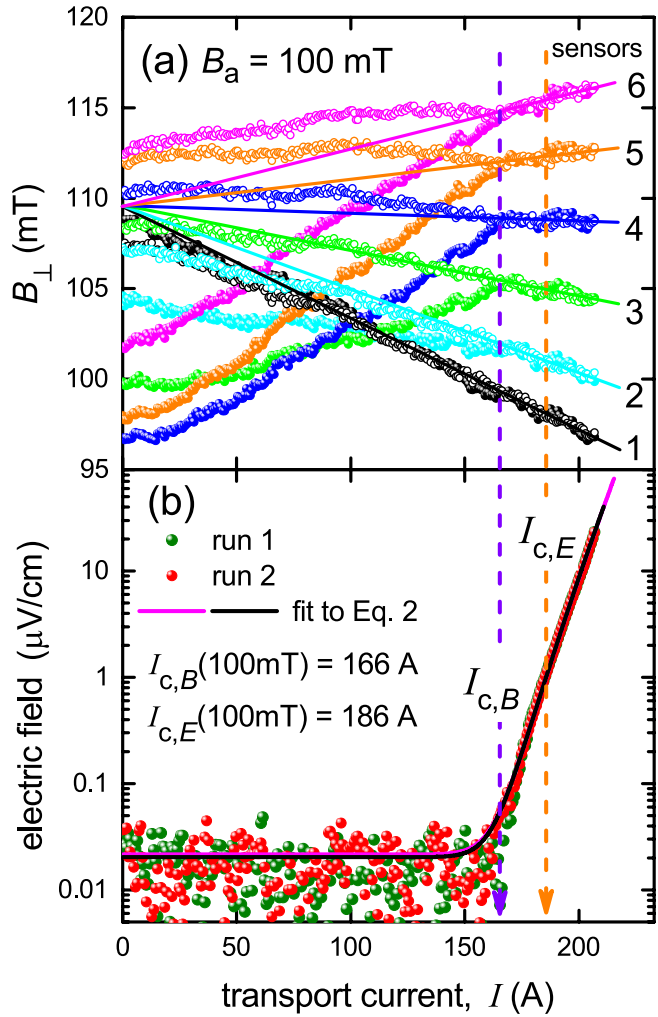

**Figure 10.** Transport current dependence of (**a**) $B_\perp(I)$ and (**b**) $E(I)$ for two consecutive ramps of transport current through a 4 mm wide SuNAM 2 G wire ($d = -0.5$ mm) at $B_a = 100$ mT. For (**a**) the filled symbols represent the first ramp, empty symbols represent the second. The solid lines show the extrapolation of the linear part of $B_\perp(I)$ back to $I = 0$ A. These lines all converge on $B_\perp(0) = 109.6 \pm 0.5$ mT $\pm$, showing the effect of the paramagnetic Hastelloy substrate in enhancing the local field. The curves $B_\perp(I)$ cross over abruptly from non-linear to linear at $I_{c,B} = 166$ A, precisely where the first observable onset of dissipation occurs. For (**b**) the fitted $n$-value is 29.5.

$$B_\perp(x_1, y_1) = B_a^* + \frac{\mu_0 I_{c,B}}{8\pi a} \ln\left\{\frac{[(a+x_1)^2 + y_1^2][(a-x_1)^2 + y_1^2]}{[(b+x_1)^2 + y_1^2][(b-x_1)^2 + y_1^2]}\right\} \quad (6)$$

Here the effective applied field, $B_a^*$, is enhanced by the susceptibility of the Hastelloy substrate[28] so that $B_a^* = \mu H_a = \mu_0(1 + \chi_v)H_a$. We find here that $B_a^* = 110 \pm 0.5$ mT, i.e. $\chi_v = 0.10 \pm 0.005$. Now the magnitude of $b$ is given by $b = 2a \exp(-B_a/B_f)$ where $B_f = \mu_0 I_{c,B}/(2\pi a)$ is the characteristic field for film geometry[15] and inserting the numbers we find $b = 0.0098$ mm. We may therefore accept our approximation noted above and, indeed, $b$ in Eq. (6) may be set to zero.

The blue curves in Fig. 9 show $B_\perp$ calculated in this way. Note that these curves modulate about the value $B_a^* = 110$ mT and not about $B_a = 100$ mT (the applied field in the absence of the substrate). The solid curve is for our effective sensor array separation of $y_1 = 0.9$ mm, established by the above zero-field calculations (and consistent with the experimental geometry). This gives quite a good account of the data though a slightly better match is found for $y_1 = 1.0$ mm. More importantly, it can be seen that the apparent lift in $B_\perp$ when the current is raised to saturation then decreased to $I = 0^+$ arises simply because of the deep trough at the tape centre in zero virgin current. The $B_\perp(I = 0^+)$ data is actually now located asymmetrically around the value of the inferred effective field, $B_a^* = 110 \pm 0.5$ mT. The origins of this asymmetry can be seen by calculating the effect of raising the current above $I_{c,B} = 166$ A. Again we could use the complex equations developed by Zeldov *et al.*[15], but the saturation of the current at 166 A means that Eq. (4) is again exactly integrable and we obtain:

$$B_\perp(x_1, y_1) = B_a^* + \frac{\mu_0 I_{c,B}}{8\pi a} \ln\left\{\frac{[(a+x_1)^2 + y_1^2]}{[(a-x_1)^2 + y_1^2]}\right\}$$ (7)

This is shown in Fig. 9 by the olive green coloured dash-dot curve which shows that, where $B_\perp(x)$ was an even function of $x_1$ in a virgin field and zero current, it becomes an odd function of $x_1$ under an in-field saturation transport current. Finally, on reducing the current back to zero, the amplitude of the profile reduces (as seen in Fig. 3) and remains an odd function of $x_1$ despite the presence of the external field. We may calculate this profile for these red data points using Eq. (48) of Zeldov et al.[15], and this is shown by the red curves in Fig. 9. In particular the calculation is done for the inferred value $B_a^* = 110$ mT with $I_{c,B} = 166$ A and $I$ raised to $I_0 = 166$ A then reduced back to $I = 0^+$. Curves calculated in the same way for $y_1 = 0.7$, 0.9 and 1.0 mm are shown and certainly $y_1 = 0.9$ mm (solid curves) provides a very satisfactory account of the detailed profile hysteresis. We wish now to shift the focus back again to the onset of dissipation but this time in the presence of both transport current and external field.

*In-field measurements – current hysteresis and dissipation onset.* In Fig. 9 we show $B_\perp(I)$ curves and $E(I)$ curves for two consecutive ramps of transport current from 0 to 210 A through 4 mm SuNAM 2 G wire under a constant applied field of $B_a = 100$ mT.

Several key features are evident:

1. Although the $B_\perp(I)$ curves for the sensors differ markedly between runs for $I < I_{c,B}$, there is an abrupt transition from non-linear to linear $B_\perp(I)$ at $I_{c,B} = 166$ A, from which point the curves coincide and are no longer hysteretic. As noted, this value was used in the calculations summarized above in Fig. 9. Similar results were found at 800 mT (not shown).
2. Extrapolation of this linear behaviour back to $I = 0$ intersects the vertical axis at $B = 109.6 \pm 0.5$ mT for all six sensors (solid lines in Fig. 10). This is essentially the same value, $B_a^* = 110 \pm 0.5$ mT, deduced from the fits in Fig. 9 and, as noted, it lies above the applied field of 100 mT due to the paramagnetic susceptibility of the Hastelloy substrate. (Presumably, for a substrate with negligible magnetic susceptibility the linear $B_\perp(I)$ data would extrapolate back to 100 mT.) It is notable that the two methods of identifying the magnitude of $B_a^*$ are independent, one based on fitting the linear range of $B_\perp(I)$ for all $x_1$ values, and the other based on the non-linear region.
3. Precisely as reported above for zero external field, panel (b) shows that the first observable onset of in-field transport dissipation (as indicated by the appearance of a measureable electric field) occurs simultaneously with the abrupt transition in $B_\perp(I)$ to linear behaviour. Note that the onset of a measureable electric field is abrupt at $I_{c,B}$. It does not follow the smooth power-law onset that is widely presumed (see fits to Eq. (1) in panel (b)). As noted, similar results were found at 800 mT, where $I_{c,B} = 19.4$ A. In both cases $I_{c,B}$ lies well below the conventionally defined $I_{c,E}$, and the electric field at $I_{c,B}$ lies nearly two orders of magnitude below that at $I_{c,E}$. This illustrates the arbitrary nature of $I_{c,E}$, contrasting with the fundamental character of $I_{c,B}$.

Very similar results were found in an applied field of 80 mT with the Fujikura tapes. We conclude that $I_{c,B}$, where $B_\perp(I)$ abruptly crosses over to linear behaviour, remains the fundamental critical current irrespective of the presence or otherwise of an externally applied magnetic field. Importantly, our calculations confirm that this coincides with the critical state abruptly reaching the centre of the tape. It does not, as it might appear, reflect a transition occurring simultaneously across the entire tape but rather, at any current, parts of the conductor, beginning at the edges, have already transitioned to the critical state and this region sweeps inwards increasingly rapidly with increasing current, becoming singular at $I_{c,B}$. Because $B_\perp(I)$ at any point on the surface represents an integral over the entire conductor (Eq. (4)) this transition is discontinuous at every point even if that point on the conductor is already in the critical state.

We note that in Fig. 10(b) the power law fit using Eq. (1) is somewhat misleading. The scatter at low current reflects fluctuations about the instrumental resolution, including negative values which cannot be displayed on the logarithmic scale. In any case we believe a pure power law fit may not be appropriate to the underlying physics. All the indications are that the onset of a dissipative voltage is abrupt and any observed smooth onset in practical conductors will rather reflect inhomogeneity in conductor performance or local variations in superfluid density[22]. For a uniform superconductor the dissipation onset is indicated as being discontinuously abrupt.

## Conclusions
We have measured, analyzed and calculated the surface perpendicular magnetic flux density generated by transport currents in second generation HTS wires, both with and without an externally applied magnetic field. We confirmed that our previous self-field $I_{c,B}$ definition of critical current is also applicable in-field. In all cases we observe that there is a distinct threshold value for transport current at which dissipation sets in. At this threshold current, $I_{c,B}$, at least three simultaneous processes occur:

1. an abrupt transition from a non-linear to a linear perpendicular field, $B_\perp(I)$, observed at all points on the superconductor surface;
2. an abrupt onset of a non-zero electric field; and
3. the transport current becomes uniformly distributed across the tape width.

At $I_{c,B}$ the entire conductor is in the critical state, the current density is uniform across the tape and any further increase in current inevitably results in (i) the onset of dissipation and (ii) a linear response in any continuing $B_\perp$ versus $I$. We stress that as such the critical current in a uniform conductor is not initiated by the nucleation of localised dissipation hotspots in weak-linked or high current density sites. Because of this we propose to designate $I_{c,B}$ as the true, fundamental critical current – one which is free from arbitrary criterion-based definitions such as customized electric field, resistance or power dissipation criteria.

In the absence of an externally applied field, the linear response of $B_\perp(I)$ extrapolates back to zero at $I=0$ but in an applied perpendicular field, $B_a$, the linear response extrapolates back to $B = \mu_0(1+\chi_v)H_a$ where $\chi_v$ is the volume magnetic susceptibility of the substrate. We found that the sharp features of $B_\perp(x_1)$ both at the edges and, in an external field, at the centre of the conductor were somewhat broadened and this was found to be consistent with the inferred physical separation between the active region of the Hall sensor array and the superconductor surface.

Finally, we stress that the detailed microscopic origins of the above-noted critical state and the associated value of $I_{c,B}$, whether driven by depinning, depairing, or limited by quantum rules remain to be fully established.

## Data Availability

The data that support the findings of this study are available from the corresponding author (EFT) upon reasonable request.

## Acknowledgements

The authors thank G.N. Sidorov for help in developing the Hall array sample mount E.F.T. thanks financial support provided by the state assignment of FASO of Russia (theme "Pressure" No. AAAA-A18-118020190104-3) and by Act 211 Government of the Russian Federation, contract No. 02.A03.21.0006.

### Author Contributions
E.F.T., J.L.T. and N.M.S. conceived the work; N.M.S. and S.C.W. designed and built the measurement system used to study the SuNAM tapes; A.E.P. and R.A.B. designed and built the measurement system used to study the Fujikura tapes; E.F.T. and J.B. performed the experiments; E.F.T., J.B. and J.L.T. analyzed the experimental data; E.F.T. revealed the presence of $B_{c1}$ front across tape width (showed in Fig. 2), J.G.S. and J.L.T. calculated the magnetic field using the Brandt-Indenbom and Zeldov-Clem-McElfresh-Darwin models or the equations presented here; E.F.T. drafted the manuscript which was revised by S.C.W. and J.L.T.

### Additional Information
**Competing Interests:** The authors declare no competing interests.

**Publisher's note:** Springer Nature remains neutral with regard to jurisdictional claims in published maps and institutional affiliations.